\RequirePackage[l2tabu, orthodox]{nag}
\documentclass[
aps,
prl,
twocolumn,
showpacs
amsmath,
amssymb,
floatfix,
longbibliography,
letterpaper,
lengthcheck,
superscriptaddress
]{revtex4}

\usepackage{graphicx}
\usepackage{rotating}
\usepackage{bbm}
\usepackage{subfigure}
\usepackage{color}
\usepackage{dsfont}
\usepackage{ulem}
\usepackage{bbold}                                                                      
\usepackage{color}
\usepackage{float}
\usepackage{hyperref}
\usepackage[all,warning]{onlyamsmath}
\usepackage[capitalise]{cleveref}                                                       

\usepackage{xstring} \noexpandarg       
\usepackage{xparse}                     

\usepackage[english]{babel}
\usepackage{amsfonts}
\usepackage{wasysym}
\usepackage{braket}
\newcommand{\sgn}{\operatorname{sgn}}

\graphicspath{
  {figures/}
}

\begin{document}


\newcommand \diff {\mathrm{d}}
\newcommand \imagi {\mathrm{i}}								
\newcommand \br {\mathbf{r}}									

\def\bra#1{\langle #1 \vert}
\def\ket#1{| #1 \rangle}
\def\braket#1#2{\langle #1 | #2 \rangle}

\newcommand{\Jrem}[1]{\textcolor{blue}{\sout{#1}}}
\newcommand{\Jadd}[1]{\textcolor{blue}{\uline{#1}}}

\newcommand \blue[1]{{\color{blue} #1}}
\newcommand \red[1]{{\color{red} #1}}


\title{Optimized Effective Potential for Quantum Electrodynamical Time-Dependent Density Functional Theory}

\author{Camilla Pellegrini}
\email[]{camilla.pellegrini@ehu.es}
\affiliation{Nano-bio Spectroscopy Group and ETSF Scientific Development Centre, Departamento de Fisica de Materiales, Universidad del Pais Vasco UPV/EHU, E-20018 San Sebastian, Spain}
\author{Johannes Flick}
\email[]{flick@fhi-berlin.mpg.de}
\affiliation{Fritz-Haber-Institut der Max-Planck-Gesellschaft, Faradayweg 4-6, D-14195 Berlin, Germany}
\author{Ilya V. Tokatly}
\email[]{ilya.tokatly@ehu.es}
\affiliation{Nano-bio Spectroscopy Group and ETSF Scientific Development Centre, Departamento de Fisica de Materiales, Universidad del Pais Vasco UPV/EHU, E-20018 San Sebastian, Spain}
\affiliation{IKERBASQUE, Basque Foundation for Science, 48011 Bilbao, Spain}
\author{Heiko Appel}
\email[]{appel@fhi-berlin.mpg.de}
\affiliation{Fritz-Haber-Institut der Max-Planck-Gesellschaft, Faradayweg 4-6, D-14195 Berlin, Germany}
\affiliation{Max Planck Institute for the Structure and Dynamics of Matter, Luruper Chaussee 149, 22761 Hamburg, Germany}
\author{Angel Rubio}
\email[]{angel.rubio@ehu.es}
\affiliation{Nano-bio Spectroscopy Group and ETSF Scientific Development Centre, Departamento de Fisica de Materiales, Universidad del Pais Vasco UPV/EHU, E-20018 San Sebastian, Spain}
\affiliation{Fritz-Haber-Institut der Max-Planck-Gesellschaft, Faradayweg 4-6, D-14195 Berlin, Germany}
\affiliation{Max Planck Institute for the Structure and Dynamics of Matter, Luruper Chaussee 149, 22761 Hamburg, Germany}

\date{\today}

\begin{abstract}
We propose a practical approximation to the exchange-correlation functional of (time-dependent) density functional theory for many-electron systems coupled to photons. The (time non-local) optimized effective potential (OEP) equation for the electron-photon system is derived. We test the new approximation in the Rabi model from weak to strong coupling regimes. It is shown that the OEP (i) improves the classical description, (ii) reproduces the quantitative behavior of the exact ground-state properties and (iii) accurately captures the dynamics entering the ultra-strong coupling regime. The present formalism opens the path to a first-principles description of correlated electron-photon systems, bridging the gap between electronic structure methods and quantum optics for real material applications. 
\end{abstract}


\maketitle

The last two decades have witnessed increasing experimental interest in the study and control of many-electron systems strongly interacting with quantum electromagnetic fields. This includes notable experiments in the areas of cavity \cite{Niemczyk2010} and circuit \cite{Houck2012} quantum electrodynamics (QED), quantum computing via photon-mediated atom entanglement \cite{vanLoo2013}, electromagnetically induced transparency \cite{K.-J.Boller1991}, quantum plasmonics \cite{Tame2013}, quantum simulators \cite{Barreiro2011} and chemistry \cite{Fontcuberta2012,Hutchison2012}. The description of realistic coupled matter-photon systems requires combining electronic structure methods from materials science with quantum-optical models. 
Recently, we have developed a time-dependent density-functional theory (TDDFT) for QED \cite{Tokatly2013,Ruggenthaler2014,Farzanehpour2014,Ruggenthaler2011} allowing for such a first-principles treatment. 
However, any application of this theory requires approximations to 
the electron-photon exchange-correlation (xc) functional, which are currently not available.

In this Letter we construct the first approximation to the xc-functional of QED-(TD)DFT. To achieve this goal, we extend the optimized effective potential (OEP) approach \cite{Ullrich1995a,Ullrich1995b,Leeuwen1996,Gorling1997,Kummel2003,Kummel2008} to the electron-photon system. The new functional is tested from the weak to the ultra-strong coupling regime in 
the Rabi model \cite{Shore1993,Braak2011,Wolf2013}, through comparison with the exact and classical solutions. We also address the functional dependence on the initial many-body state, assumed to be either a fully interacting or a factorizable state. In both cases, the electron-photon OEP performs well, providing a promising path for describing complex strongly coupled matter-photon systems.

Consider a system with an arbitrarily large number $N$ of electrons at coordinates $\{\textbf{r}_i\}_{i=1}^{N}$, e.g. an atom, an ion, or a molecular cluster, interacting with $M$ quantized radiation modes of momenta $\{p_{\alpha}\}_{\alpha=1}^{M}$ and frequencies $\omega_{\alpha}$. We denote by $\hat{H}_0=\hat{T}+\hat{V}_{\textup{ee}}+\hat{V}_{\textup{ext}}$ the Hamiltonian of the unperturbed electronic system with kinetic energy $\hat{T}$, Coulomb interaction $\hat{V}_{\textup{ee}}$, and external (time-dependent) potential $\hat{V}_{\textup{ext}}=\sum_{i=1}^N v_{\textup{ext}}(\textbf{r}_it)$, due to the nuclear attraction and any classical field applied to the system. Adopting the length gauge in the usual dipole approximation \footnote{The derivation can be generalized to the case of atom-field coupling beyond the dipole-approximation in straightforward manner.}, the Hamiltonian of the fully interacting electron-photon system takes the form \cite{Tokatly2013,Faisal}
\begin{equation}
\label{eqn:H}
\hat{H}=\hat{H}_0+\frac{1}{2}\sum_{\alpha}\left[\hat{p}^2_{\alpha}+\omega^2_{\alpha}\left(\hat{q}_{\alpha}-\frac{\boldsymbol{\lambda}_{\alpha}}{\omega_{\alpha}}\hat{\textbf{R}}\right)^2\right],
\end{equation}
where $\hat{\textbf{R}}\!=\!\sum_{i=1}^{\textbf{N}}\textbf{r}_i$ is the dipole moment operator of the electronic system and $\boldsymbol{\lambda}_{\alpha}$ is the coupling constant of the $\alpha$ photon mode. We define the photon coordinate in terms of annihilation and creation operators as $\hat{q}_{\alpha}\!=\!-(\hat{a}_{\alpha}+\hat{a}^\dag_{\alpha})/\sqrt{2\omega_{\alpha}}$.
The interaction Hamiltonian in Eq.~\eqref{eqn:H} consists of two terms. The cross term $\sim\hat{q}_{\alpha}\hat{R}$, i.e. 
\begin{equation}
\label{eqn:Velph}
\hat{V}_{\textup{el-ph}}=\sum_{\alpha}\sqrt{\frac{\omega_{\alpha}}{2}}(\hat{a}_{\alpha}+\hat{a}^{\dag}_{\alpha})\int d^3 \textbf{r}\left(\boldsymbol{\lambda}_{\alpha} \textbf{r}\right)\hat{n}(\textbf{r}),
\end{equation}
describes the dipole-photon coupling, where $\hat{n}(\textbf{r})=\sum_i\delta(\textbf{r}-\textbf{r}_i)$ is the electron density operator. The squared term $\sum_{\alpha}(\boldsymbol{\lambda}_{\alpha}\textbf{R})^2/2$ represents the polarization energy of the electronic system due to the Coulomb-like interaction $\hat{v}(\textbf{r}_i,\textbf{r}_j)=\sum_{\alpha}(\boldsymbol{\lambda}_{\alpha}\textbf{r}_i)(\boldsymbol{\lambda}_{\alpha}\textbf{r}_j)$. It follows that the electron-photon coupling gives rise to an additional electron-electron interaction of the form 
\begin{align}
\label{eqn:int}
W_{\textup{ee}}(1,2)&=\sum_{\alpha}(\boldsymbol{\lambda}_{\alpha}\textbf{r}_1)(\boldsymbol{\lambda}_{\alpha}\textbf{r}_2)\mathcal{W}(t_1,t_2),\\
&\mathcal{W}(t_1,t_2)=\omega^2_{\alpha}D(t_1,t_2)+\delta(t_1-t_2),\nonumber
\end{align}
where we use the compact notation $1=(\textbf{r}_1t_1)$.
Here the first term, which involves the photon propagator $\textup{i}D(t_1,t_2)\!\equiv\!\langle\mathcal{T}\,\{q_{\alpha}(t_1)\,q_{\alpha}(t_2)\}\rangle$, corresponds to the effective photon-mediated interaction derived from Eq.~\eqref{eqn:Velph}. The second term accounts for the electrostatic interaction $v$. We use atomic units (a.u.) throughout the paper.

Our formulation of QED-(TD)DFT combines one of the most popular exact methods for ground (excited) state electronic calculations \cite{Marques}, with the full quantum treatment of the electromagnetic field. In this theory, the wave function of the total system $\Psi(\{\textbf{r}_j\},\{q_{\alpha}\},t)$ is a unique functional of the electron density $n(\textbf{r}t)=\bra{\Psi}\hat{n}(\textbf{r})\ket{\Psi}$ and the expectation values of the photon coordinates $Q_{\alpha}(t)=\bra{\Psi}\hat{q}_{\alpha}\ket{\Psi}$ \cite{Tokatly2013}. The former can be calculated for a fictitious Kohn-Sham (KS) system of $N$ non-interacting particles, whose orbitals $\{\phi_j\}$ satisfy the self-consistent equations $\textup{i}\partial_t\phi_j(\textbf{r}t)=\left[-\nabla^2/2+v_s(\textbf{r}t)\right]\phi_j(\textbf{r}t)$ with the potential $v_{\textup{s}}=v_{\textup{ext}}+v_{\textup{eff}}$. Here the effective part of the KS potential is defined as $v_{\textup{eff}}=v_{\textup{MF}}+v_{\textup{xc}}$, where
\begin{equation}
\label{eqn:vh}
v_\textup{MF}(\textbf{r}t)=\int d1\, W^{R}_{\textup{ee}}(\textbf{r}t,\textbf{r}_1t_1)\,n(\textbf{r}_1t_1)
\end{equation}
is the mean-field contribution due to $M$ classical electromagnetic modes, whose expectation values $Q_{\alpha}$ obey the Ampere-Maxwell equation for the electric field. All the quantum many-body effects are embedded in the unknown xc-potential $v_{\textup{xc}}$, which must be approximated. In the following we disregard the contribution of the Coulomb interaction $V_{\textup{ee}}$, which can be treated by standard methods.

In this work we generalize the OEP approach to construct approximations to the electron-photon functional $v_{\textup{xc}}$ \footnote{In contrast to the usual OEP scheme for the treatment of the Coulomb interaction $V_{\textup{ee}}$ \cite{Ullrich1995a,Kummel2003}, we perform the approximation in the coupling $W_{\textup{ee}}$ to transversal photon modes.}.
We derive the TDOEP equation for the electron-photon system in the form of the generalized linear Sham-Schl\"{u}ter equation on the Keldysh contour \cite{Leeuwen1996}
\begin{equation}
\begin{aligned}
\label{eqn:OEP}
\!\!\!\int\!\!\! d2 G_{\textup{s}}(1,\!2)v_{\textup{xc}}(2)G_{\textup{s}}(2,\!1)\!=\!\!\!\int\!\!\! d2 \!\!\!\int\!\!\! d3G_{\textup{s}}(1,\!2)\Sigma(2,\!3)G_{\textup{s}}(3,\!1),
\end{aligned}
\end{equation}
where the electron self-energy $\Sigma$ contains the interaction $W_{\textup{ee}}$ of Eq.~\eqref{eqn:int}.
Connecting $\Sigma$ at each order in the expansion to $v_{\textup{xc}}$, Eq.~\eqref{eqn:OEP} allows one to perturbatively construct the xc-functional to any desired order in the coupling strength $\lambda_{\alpha}$. Analogously to the $GW$ approximation \cite{Hedin1965,Onida2002} for electronic structure methods, we approximate the electron self-energy by the exchange-like diagram
\begin{equation}
\label{eqn:sigma}
\Sigma(1,2)=\textup{i}\,G_{\textup{s}}(1,2)W_{\textup{ee}}(2,1),
\end{equation}
where we assume the photon propagator in $W_{\textup{ee}}$ to be free. Here the quantum nature of the electromagnetic field is accounted for by the dynamical part of $\Sigma$, related to the first term of Eq.~\eqref{eqn:int}. This part describes the processes of emission and absorption of a photon. Neglecting the above dynamical contribution to $v_{\textup{eff}}$ corresponds to the classical treatment of the electromagnetic field.
\begin{widetext}
Making use of the Langreth rules, we obtain from Eq.~\eqref{eqn:OEP}
\begin{equation}
\label{eqn:a}
\textup{i}\int_{-\infty}^{t}\!\!\!dt_1 G^R(t,t_1) v_{\textup{x}}(t_1) G^<(t_1,t)+c.c.=
\textup{i}\int_{-\infty}^{t}\!\!\!dt_1\!\int_{-\infty}^{t_1}\!\!\!dt_2\, G^R(t,t_1)[\Sigma^>(t_1,t_2)G^<(t_2,t)-\Sigma^<(t_1,t_2)G^>(t_2,t)]+c.c.,
\end{equation}
\end{widetext}
where the integration over the spatial coordinates is implied. For computational convenience we consider Eq.~\eqref{eqn:a} in the low temperature limit $T\rightarrow0$. The electron-photon collision integral on the right hand side then is responsible for the spontaneous photon emission of the excited electrons and the broadening in the electronic levels. Eq.~\eqref{eqn:a} explicitly reads as
\begin{align}
\label{eqn:TDOEP}
&\textup{i}\sum_{i,j}\!\int_{-\infty}^{t}\!\!\!\!\!\!dt_1\!\left[\bra{\phi_i(t_1)}v_{\textup{x}}(t_1)\ket{\phi_j(t_1)}f_i-S_{ij}(t_1)\right]\phi^*_j(t)\phi_i(t)\nonumber\\
&+c.c.=0,
\end{align}
where we define
\begin{equation}
\begin{aligned}
S_{ij}(t_1)\!=&\!\sum_{k,\alpha}\int_{-\infty}^{t_1}\!\!\!\!\!\!dt_2\, d^{\alpha}_{ik}(t_2)d^{\alpha}_{kj}(t_1)[(1-f_i)f_k \mathcal{W}^>(t_1,t_2)\nonumber\\
&-f_i(1-f_k)\mathcal{W}^<(t_1,t_2)]
\end{aligned}
\end{equation}
with $\mathcal{W}^\gtrless(t_1,t_2)=\omega^2_{\alpha}\Big(\frac{-\textup{i}}{2\omega_{\alpha}}\Big)e^{\pm \textup{i}\omega_{\alpha}(t_2-t_1)}\pm \delta(t_1-t_2).\\$ Here $f_i$ is the fermion occupation number and $d^{\alpha}_{ik}(t)=\boldsymbol{\lambda}_{\alpha}\bra{\phi_i(t)}\textbf{r}\ket{\phi_k(t)}$ is the projection of the dipole matrix element on the coupling constant of the $\alpha$-mode. The orbitals $\{\phi_j\}$ are solutions of the time-dependent KS equations with the initial condition $\phi_j(\textbf{r}t)=\phi_j(\textbf{r})e^{-\textup{i}\varepsilon_jt}$ for $-\infty<t\leq0$. Alternatively, Eq.~\eqref{eqn:TDOEP} can be derived via the variational principle from the action functional on the Keldysh contour, with the exchange part given by
\begin{align*}
A_{\textup{x}}\!=&\!\sum_{i,k,\alpha}\int\!\! dz_1 \!\!\!\int\!\! dz_2\,
d^{\alpha}_{ik}(z_2)d^{\alpha}_{ki}(z_1)(1-f_i)f_k\\
&\times\theta(z_1-z_2)\left[\omega^2_{\alpha}\Big(\frac{-\textup{i}}{2\omega_{\alpha}}\Big)e^{\textup{i}\omega_{\alpha}(z_2-z_1)}+\delta(z_1-z_2)\right],\nonumber
\end{align*}
where we denote the contour variable by $z$.
Furthermore, the time-dependent mean-field potential is evaluated from Eq.~\eqref{eqn:vh} as
\begin{align}
\label{eqn:VH}
&v_{\textup{MF}}(\textbf{r}t)\!=\!-\!\!\sum_{\alpha}\omega_{\alpha}(\boldsymbol{\lambda}_{\alpha}\textbf{r})\!\!\int_0^{t} \!\!\!dt_1\sin\!\left[\omega_{\alpha}(t\!-\!t_1)\right]\!(\boldsymbol{\lambda}_{\alpha}\textbf{R}(t_1))\nonumber\\
&-\!\!\sum_{\alpha}(\boldsymbol{\lambda}_{\alpha}\textbf{r})\left[(\boldsymbol{\lambda}_{\alpha}\textbf{R}(0))\cos(\omega_{\alpha}t)-(\boldsymbol{\lambda}_{\alpha}\textbf{R}(t))\right],
\end{align}
where $\textbf{R}(t)=\int d^3\textbf{r}\,\textbf{r}\, n(\textbf{r}t)$ is the expectation value of the dipole moment operator of the electronic system.
\begin{figure}[t]
   \centering \includegraphics[height=0.27\textheight]{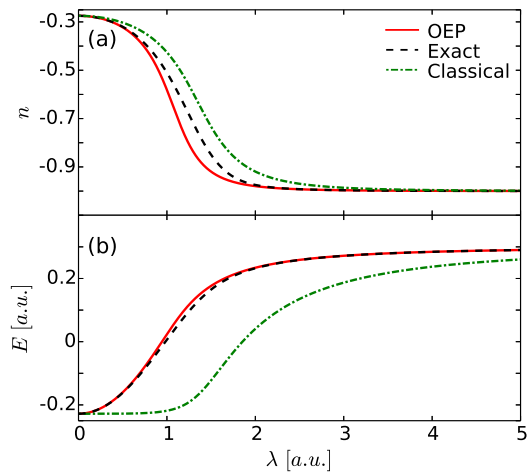}
  \caption{(Color online) Comparison of the OEP (red), exact (black) and classical (green) (a) ground-state density $n$ and (b) energy $E$ versus the coupling parameter $\lambda$ in a.u.. Other parameters: $\omega=1$, $v_{\textup{ext}}=0.2$, $t_{\textup{kin}}=0.7$.}
  \label{fig:static}
\end{figure}
In the special case of the electron-photon system in equilibrium at time $t=0$ with $V_{\textup{ext}}=V_{\textup{ext}}(0)$, Eq.~\eqref{eqn:TDOEP} reduces to the stationary OEP equation
\begin{equation}
\label{eqn:staticOEP}
\sum_{i,j}\left[\frac{\bra{\phi_i}v_{\textup{x}}\ket{\phi_j}}{\varepsilon_i\!-\!\varepsilon_j\!-\textup{i}\eta}f_i-S_{ij}\right]\phi^*_j(\textbf{r})\phi_i(\textbf{r})+c.c.=0,
\end{equation}
where
\begin{align}
\label{eqn:S}
S_{ij}=&\sum_{k,\alpha}\frac{d^{\alpha}_{ik}d^{\alpha}_{kj}(\varepsilon_i-\varepsilon_k-i\eta)}{2(\varepsilon_i-\varepsilon_j-\textup{i}\eta)}\Big[\frac{f_i(1-f_k)}{\varepsilon_i-\varepsilon_k-\omega_{\alpha}-\textup{i}\eta}\nonumber\\
&+\frac{(1-f_i)f_k}{\varepsilon_i-\varepsilon_k+\omega_{\alpha}-\textup{i}\eta}\Big].
\end{align}
Here we assume the limit $\eta\rightarrow0$.
Eq.~\eqref{eqn:S} describes the virtual process of excitation of electron-hole pairs, supplemented with the virtual emission of a photon. Eq.~\eqref{eqn:staticOEP} can be variationally derived employing the second-order correction to the ground-state energy
\begin{equation}
\label{eqn:Exc}
E_{\textup{x}}= -\frac{1}{2}\sum_{i,k,\alpha} |d^{\alpha}_{ik}|^2\,\Big\{ \omega_{\alpha}\,\frac{(1-f_i)f_k}{\varepsilon_i-\varepsilon_k+\omega_{\alpha}}-(1-f_i)f_k \Big\}.
\end{equation}
The first term in the x-energy $E_{\textup{x}}$ is interpreted as the Lamb shift due to the virtual emission of photons. The second term comes from the counter-term $\sum_{\alpha}(\boldsymbol{\lambda}_{\alpha}\textbf{R})^2/2$ in the Hamiltonian, and accounts for the free electron behavior in the high photon energy limit $\omega_{\alpha}\rightarrow\infty$.

We now apply these results to a simple exactly solvable cavity-QED system, i.e.\ the two-level model of a diatomic molecule with one electron in a single-mode cavity. For the one photon case, and in the one-electron basis set ($\hat{\textbf{r}}\!\rightarrow\!\hat{\sigma}_z, \hat{T}\!\rightarrow\!-t_{\textup{kin}}\hat{\sigma}_x$), the Hamiltonian $\hat{H}$ of Eq.~\eqref{eqn:H}
\begin{equation}
\label{eqn:H1}
\hat{H}=-t_{\textup{kin}}\hat{\sigma}_x+[g(\hat{a}+\hat{a}^\dag)+v_{\textup{ext}}(t)] \hat{\sigma}_z+\omega\Big(\hat{a}^\dag \hat{a}+\frac{1}{2}\Big)+\frac{\lambda^2}{2}
\end{equation}
is isomorphic to the generalized Rabi Hamiltonian with external potential $v_{\textup{ext}}(t)$ and coupling strength $g=\sqrt{\omega/2}\,\lambda$.

We consider first the description of the system in equilibrium at time $t\!=\!0$ with $v_{\textup{ext}}=v_{\textup{ext}}(0)$.
Eq.~\eqref{eqn:staticOEP}, projected onto $\phi^\dag_1\!=\!(\bar{v}\,\,\,\bar{u})$ and $\phi^\dag_2\!=\!(\bar{u}\,\, -\bar{v})$, with related eigenvalues $\varepsilon_1=-W$ and $\varepsilon_2=W$,
where $\bar{u},\bar{v}=\sqrt{\left(1\pm v_{\textup{s}}/W\right)/2}$ and $W=\sqrt{v_{\textup{s}}^2+t^2_{\textup{kin}}}$, gives
\begin{equation}
\label{eqn:explicitVS}
v_{\textup{x}}=-\lambda^2\frac{v_{\textup{s}}}{W}\left[\,\frac{\omega(\omega+3W)}{(\omega+2W)^2}-1\right].
\end{equation}
Here the second term corresponds to the classical contribution associated with the first interaction term in Eq.~\eqref{eqn:int}.
Moreover, the total energy of the system takes the form
\begin{equation}
\label{eqn:EK}
\begin{split}
E[v_{\textup{s}}]\!=\!-t_{\textup{kin}} \langle \sigma_x \rangle\!+\!v_{\textup{ext}}n\!+\!E_{\textup{x}}[v_{\textup{s}}]\!+\!\frac{1}{2}\omega,
\end{split}
\end{equation}
where $n=\!\langle\sigma_z\rangle\!=\!-v_{\textup{s}}/W$ is the density difference between the two levels, and Eq.~\eqref{eqn:Exc} reduces to
\begin{equation}
\label{eqn:Ex}
E_{\textup{x}}=\frac{\lambda^2t^2_{\textup{kin}}}{W\,(\omega+2\,W)}.
\end{equation}
The net exchange contribution to the ground-state total energy vanishes in the classical limit of coupling $\lambda\rightarrow\infty$, as expected. 
In Fig.~\ref{fig:static} we show the calculated OEP ground-state density $n$ and total energy $E$ as functions of the coupling strength $\lambda$, compared to the results obtained from the exact and classical treatment of the electromagnetic field. The eigenvalue problem for the static Rabi Hamiltonian in Eq.~\eqref{eqn:H1} is solved by employing the exact diagonalization technique \cite{Streltsov2010,Flick2014}, after proper truncation of the Fock space. 
We observe that both the OEP and classical approximations reproduce qualitatively the electron's confinement on the excited level, as the shift in the energy levels increases with the coupling strength, and recover the exact result in the limit $\lambda\rightarrow\infty$. 
In addition, our OEP scheme is by construction exact in the weak coupling regime.
For the ground-state densities $n$ shown in (a), we see excellent agreement between the OEP and the exact results up to $\lambda = 0.5$ and above $\lambda = 2$. In contrast, the classical approximation performs reliably only in the limits of very small or very high interaction strength. Regarding the ground-state total energies $E$ shown in (b), the improvement of the OEP with respect to the classical approach is evident. Here the classical result is only asymptotically accurate and largely underestimating in between. On the contrary, the OEP energy is close to the exact values in the whole coupling range, with only small deviations around $\lambda = 1.25$.

The TDOEP Eq.~\eqref{eqn:TDOEP} for the Rabi model simplifies to
\begin{align}
\label{eqn:TDOEP-rabi}
&\textup{i}\int_{-\infty}^t dt_1\tilde{v}_{\textup{x}}(t_1)d_{12}(t_1)d_{21}(t)+c.c.\nonumber\\
&=\lambda^2\omega\int_{-\infty}^t\!\!\!\! dt_1\int_{-\infty}^{t_1}\!\!\!\! dt_2\,c(t,t_1)\,d_{21}(t_2)e^{\textup{i}\omega(t_2-t_1)}+c.c.,
\end{align}
where $\tilde{v}_{\textup{x}}=v_{\textup{x}}(t)+\lambda^2n(t)$ and $c(t,t_1)=d_{12}(t)n(t_1)-d_{12}(t_1)n(t)$. 
Moreover, the mean-field potential of Eq.~\eqref{eqn:VH} explicitly reads as
\begin{equation}
\begin{aligned}
\label{eqn:VHR}
v_{\textup{MF}}(t)\!=&\!-\lambda^2\omega\!\int_0^{t} \!\!\!dt_1\sin\!\left[\omega(t\!-\!t_1)\right]\!n(t_1)\!-\!\lambda^2n \cos(\omega t)\nonumber\\
&+\lambda^2n(t).
\end{aligned}
\end{equation}
\begin{figure}[t]
  \centering \includegraphics[width=0.49\textwidth]{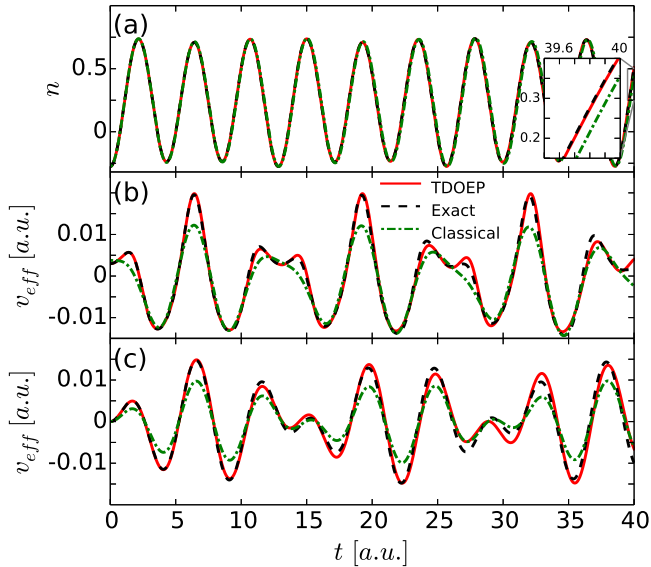}
  \caption{(Color online) Comparison of the TDOEP (red), exact (black) and classical (green) (a) density $n$ and (b, c) effective potential $v_{\textup{eff}}$ versus time $t$ in a.u. for the configurations : (a, b) $v_{\textup{ext}}\!=\!-0.2\sgn(t)$, $\lambda\!=\!0.1$ and (c) $v_{\textup{ext}}\!=\!0$, $\lambda\!=\!0.1\theta(t)$. Other parameters: $\omega\!=\!1$, $t_{\textup{kin}}\!=\!0.7$. A zoom-in of the density for $39.5\leq t\leq40$ is shown as an inset in panel (a).}
\label{fig:sudden_switch}
\end{figure}
Employing the numerical algorithm presented in \cite{Wijewardane2008}, we solve Eq.~\eqref{eqn:TDOEP-rabi} self-consistently for $t>0$, together with the time-dependent KS equation. The former, which is a Volterra integral equation of the first kind, is evaluated using a midpoint integration scheme combined with the trapezoidal rule \cite{Linz1969}. The latter is propagated with a predictor-corrector scheme using an exponential midpoint propagator \cite{Castro2004}.
In Fig.~\ref{fig:sudden_switch} we show the time-evolution of the calculated TDOEP density $n$ and effective potential $v_{\textup{eff}}$ for two different setups, compared to the exact and classical results. In the first setting, we assume that the electron-photon system, interacting with coupling constant $\lambda\!=\!0.1$, is driven out of equilibrium at $t\!=\!0$ by a sudden switch in the external perturbation $v_{\textup{ext}}(t)\! =\!-0.2\sgn(t)$. In the second configuration, we choose a non-interacting initial state with $v_{\textup{ext}}(t)\!=\!0$, while switching on at later times the electron-photon coupling $\lambda(t)\!=\!0.1\,\theta(t)$. 
Here we use as initial state for the propagation $\ket{\Psi} = (1/2 \ket{1} + \sqrt{3}/2 \ket{2})\otimes\ket{0}$, where $\ket{1}$ and $\ket{2}$ are the basis vectors of the electron system, and $\ket{0}$ is the photon vacuum field. For the chosen parameters, the various densities of the two setups undergo off-resonant Rabi oscillations with nearly identical relative behavior. This is shown in (a) for the sudden-switch example. Within the plotted range, the TDOEP and exact results are practically on top of each other. In contrast, the classical density starts to deviate around $t\!=\!20$ a.u., and the error becomes quite sizable at $t\!=\!40$ a.u.. More significant is the improvement of the TDOEP approach against the classical approximation in the effective potential. As we can see in (b) for the sudden-switch case, and in (c) for the non-interacting initial configuration, the TDOEP result is very accurate up to $t\!=\!20$ a.u.. At later times, small deviations appear, especially in (c), where the potential shows a more complex dynamics. Nevertheless, the improvement with respect to the classical result is still evident.

In conclusion, a new efficient approach for non-relativistic many-electron systems strongly interacting with photons is proposed. We showed that the lowest order (TD)OEP for the off-resonant Rabi model gives accurate ground-state properties (dynamics) far beyond the weak-coupling regime, clearly improving over the classical description. This work opens the path to a computationally efficient description of novel phenomena at the interface between condensed matter physics and quantum optics.

We acknowledge financial support by the European Research Council Advanced Grant DYNamo (ERC-2010-AdG-267374), European Commission project CRONOS (Grant number 280879-2 CRONOS CP-FP7), Spanish Grant (FIS2013-46159-C3-1-P), Ikerbasque and Grupo Consolidado UPV/EHU del Gobierno Vasco (IT578-13).

\renewcommand{\emph}[1]{\textit{#1}}


\begin{thebibliography}{28}
\expandafter\ifx\csname natexlab\endcsname\relax\def\natexlab#1{#1}\fi
\expandafter\ifx\csname bibnamefont\endcsname\relax
  \def\bibnamefont#1{#1}\fi
\expandafter\ifx\csname bibfnamefont\endcsname\relax
  \def\bibfnamefont#1{#1}\fi
\expandafter\ifx\csname citenamefont\endcsname\relax
  \def\citenamefont#1{#1}\fi
\expandafter\ifx\csname url\endcsname\relax
  \def\url#1{\texttt{#1}}\fi
\expandafter\ifx\csname urlprefix\endcsname\relax\def\urlprefix{URL }\fi
\providecommand{\bibinfo}[2]{#2}
\providecommand{\eprint}[2][]{\url{#2}}

\bibitem[{\citenamefont{Niemczyk et~al.}(2010)\citenamefont{Niemczyk, Deppe,
  Huebl, Menzel, Hocke, Schwarz, Garcia-Ripoli, Zueco, H{\"u}mmer, Solano
  et~al.}}]{Niemczyk2010}
\bibinfo{author}{\bibfnamefont{T.}~\bibnamefont{Niemczyk}},
  \bibinfo{author}{\bibfnamefont{F.}~\bibnamefont{Deppe}},
  \bibinfo{author}{\bibfnamefont{H.}~\bibnamefont{Huebl}},
  \bibinfo{author}{\bibfnamefont{E.~P.} \bibnamefont{Menzel}},
  \bibinfo{author}{\bibfnamefont{F.}~\bibnamefont{Hocke}},
  \bibinfo{author}{\bibfnamefont{M.~J.} \bibnamefont{Schwarz}},
  \bibinfo{author}{\bibfnamefont{J.~J.} \bibnamefont{Garcia-Ripoli}},
  \bibinfo{author}{\bibfnamefont{D.}~\bibnamefont{Zueco}},
  \bibinfo{author}{\bibfnamefont{T.}~\bibnamefont{H{\"u}mmer}},
  \bibinfo{author}{\bibfnamefont{E.}~\bibnamefont{Solano}}
  \bibnamefont{et~al.}, \bibinfo{journal}{Nature Physics}
  \textbf{\bibinfo{volume}{6}}, \bibinfo{pages}{772} (\bibinfo{year}{2010}).

\bibitem[{\citenamefont{Houck et~al.}(2012)\citenamefont{Houck, T{\"u}reci and
  Koch}}]{Houck2012}
\bibinfo{author}{\bibfnamefont{A.~A.} \bibnamefont{Houck}},
  \bibinfo{author}{\bibfnamefont{H.~E.} \bibnamefont{T{\"u}reci}}
  \bibnamefont{and} \bibinfo{author}{\bibfnamefont{J.}~\bibnamefont{Koch}},
  \bibinfo{journal}{Nature} \textbf{\bibinfo{volume}{8}}, \bibinfo{pages}{292}
  (\bibinfo{year}{2012}).

\bibitem[{\citenamefont{van Loo et~al.}(2013)\citenamefont{van Loo, Fedorov,
  Lalumi{\`e}re, Sanders, Blais and Wallraff}}]{vanLoo2013}
\bibinfo{author}{\bibfnamefont{A.~F.} \bibnamefont{van Loo}},
  \bibinfo{author}{\bibfnamefont{A.}~\bibnamefont{Fedorov}},
  \bibinfo{author}{\bibfnamefont{K.}~\bibnamefont{Lalumi{\`e}re}},
  \bibinfo{author}{\bibfnamefont{B.~C.} \bibnamefont{Sanders}},
  \bibinfo{author}{\bibfnamefont{A.}~\bibnamefont{Blais}} \bibnamefont{and}
  \bibinfo{author}{\bibfnamefont{A.}~\bibnamefont{Wallraff}},
  \bibinfo{journal}{Science} \textbf{\bibinfo{volume}{342 (6165)}},
  \bibinfo{pages}{1494} (\bibinfo{year}{2013}).

\bibitem[{\citenamefont{Boller et~al.}(1991)\citenamefont{Boller, Imamoglu and
  Harris}}]{K.-J.Boller1991}
\bibinfo{author}{\bibfnamefont{K.-J.} \bibnamefont{Boller}},
  \bibinfo{author}{\bibfnamefont{A.}~\bibnamefont{Imamoglu}} \bibnamefont{and}
  \bibinfo{author}{\bibfnamefont{S.~E.} \bibnamefont{Harris}},
  \bibinfo{journal}{Phys. Rev. Lett.} \textbf{\bibinfo{volume}{66}},
  \bibinfo{pages}{2593} (\bibinfo{year}{1991}).

\bibitem[{\citenamefont{Tame et~al.}(2013)\citenamefont{Tame, McEnery,
  {\"O}zdemir, Lee, Maier, and Kim}}]{Tame2013}
\bibinfo{author}{\bibfnamefont{M.~S.} \bibnamefont{Tame}},
  \bibinfo{author}{\bibfnamefont{K.~R.} \bibnamefont{McEnery}},
  \bibinfo{author}{\bibfnamefont{{\c S}.~K.} \bibnamefont{{\"O}zdemir}},
  \bibinfo{author}{\bibfnamefont{J.}~\bibnamefont{Lee}},
  \bibinfo{author}{\bibfnamefont{S.~A.} \bibnamefont{Maier}} \bibnamefont{and}
  \bibinfo{author}{\bibfnamefont{M.~S.} \bibnamefont{Kim}},
  \bibinfo{journal}{Nature Phys.} \textbf{\bibinfo{volume}{9}},
  \bibinfo{pages}{329} (\bibinfo{year}{2013}).

\bibitem[{\citenamefont{Barreiro et~al.}(2011)\citenamefont{Barreiro,
  M{\"u}ller, Schindler, Nigg, Monz, Chwalla, Hennrich, Roos, Zoller, and
  Blatt}}]{Barreiro2011}
\bibinfo{author}{\bibfnamefont{J.~T.} \bibnamefont{Barreiro}},
  \bibinfo{author}{\bibfnamefont{M.}~\bibnamefont{M{\"u}ller}},
  \bibinfo{author}{\bibfnamefont{P.}~\bibnamefont{Schindler}},
  \bibinfo{author}{\bibfnamefont{D.}~\bibnamefont{Nigg}},
  \bibinfo{author}{\bibfnamefont{T.}~\bibnamefont{Monz}},
  \bibinfo{author}{\bibfnamefont{M.}~\bibnamefont{Chwalla}},
  \bibinfo{author}{\bibfnamefont{M.}~\bibnamefont{Hennrich}},
  \bibinfo{author}{\bibfnamefont{C.~F.} \bibnamefont{Roos}},
  \bibinfo{author}{\bibfnamefont{P.}~\bibnamefont{Zoller}} \bibnamefont{and}
  \bibinfo{author}{\bibfnamefont{R.}~\bibnamefont{Blatt}},
  \bibinfo{journal}{Nature} \textbf{\bibinfo{volume}{470}},
  \bibinfo{pages}{486} (\bibinfo{year}{2011}).

\bibitem[{\citenamefont{Fontcuberta~i Morral and
  Stellacci}(2012)}]{Fontcuberta2012}
\bibinfo{author}{\bibfnamefont{A.}~\bibnamefont{Fontcuberta~i Morral}}
  \bibnamefont{and}
  \bibinfo{author}{\bibfnamefont{F.}~\bibnamefont{Stellacci}},
  \bibinfo{journal}{Nature Materials} \textbf{\bibinfo{volume}{11}},
  \bibinfo{pages}{272} (\bibinfo{year}{2012}).

\bibitem[{\citenamefont{Hutchison et~al.}(2012)\citenamefont{Hutchison,
  Schwartz, Genet, Devaux, and Ebbesen}}]{Hutchison2012}
\bibinfo{author}{\bibfnamefont{J.~A.} \bibnamefont{Hutchison}},
  \bibinfo{author}{\bibfnamefont{T.}~\bibnamefont{Schwartz}},
  \bibinfo{author}{\bibfnamefont{C.}~\bibnamefont{Genet}},
  \bibinfo{author}{\bibfnamefont{E.}~\bibnamefont{Devaux}} \bibnamefont{and}
  \bibinfo{author}{\bibfnamefont{T.~W.} \bibnamefont{Ebbesen}},
  \bibinfo{journal}{Angew. Chem. Int. Ed.} \textbf{\bibinfo{volume}{51 (7)}},
  \bibinfo{pages}{1592} (\bibinfo{year}{2012}).

\bibitem[{\citenamefont{Tokatly}(2013)}]{Tokatly2013}
\bibinfo{author}{\bibfnamefont{I.~V.} \bibnamefont{Tokatly}},
  \bibinfo{journal}{Phys. Rev. Lett.} \textbf{\bibinfo{volume}{110}},
  \bibinfo{pages}{233001} (\bibinfo{year}{2013}).

\bibitem[{\citenamefont{Ruggenthaler et~al.}(2014)\citenamefont{Ruggenthaler,
  Flick, Pellegrini, Appel, Tokatly and Rubio}}]{Ruggenthaler2014}
\bibinfo{author}{\bibfnamefont{M.}~\bibnamefont{Ruggenthaler}},
  \bibinfo{author}{\bibfnamefont{J.}~\bibnamefont{Flick}},
  \bibinfo{author}{\bibfnamefont{C.}~\bibnamefont{Pellegrini}},
  \bibinfo{author}{\bibfnamefont{H.}~\bibnamefont{Appel}},
  \bibinfo{author}{\bibfnamefont{I.~V.} \bibnamefont{Tokatly}}
  \bibnamefont{and} \bibinfo{author}{\bibfnamefont{A.}~\bibnamefont{Rubio}},
  \bibinfo{journal}{Phys. Rev. A} \textbf{\bibinfo{volume}{90}},
  \bibinfo{pages}{012508} (\bibinfo{year}{2014}).

\bibitem[{\citenamefont{Farzanehpour et~al.}(2014)\citenamefont{Farzanehpour and Tokatly}}]{Farzanehpour2014}
\bibinfo{author}{\bibfnamefont{M.}~\bibnamefont{Farzanehpour}}
  \bibnamefont{and} \bibinfo{author}{\bibfnamefont{I.~V.}~\bibnamefont{Tokatly}},
   \bibinfo{journal}{Phys. Rev. B} \textbf{\bibinfo{volume}{90}},
  \bibinfo{pages}{195149} (\bibinfo{year}{2014}).

\bibitem[{\citenamefont{Ruggenthaler et~al.}(2011)\citenamefont{Ruggenthaler,
  Mackenroth and Bauer}}]{Ruggenthaler2011}
\bibinfo{author}{\bibfnamefont{M.}~\bibnamefont{Ruggenthaler}},
  \bibinfo{author}{\bibfnamefont{F.}~\bibnamefont{Mackenroth}}
  \bibnamefont{and} \bibinfo{author}{\bibfnamefont{D.}~\bibnamefont{Bauer}},
  \bibinfo{journal}{Phys. Rev. A} \textbf{\bibinfo{volume}{84}},
  \bibinfo{pages}{042107} (\bibinfo{year}{2011}).

\bibitem[{\citenamefont{Ullrich et~al.}(1995)\citenamefont{Ullrich, Gossmann
  and Gross}}]{Ullrich1995a}
\bibinfo{author}{\bibfnamefont{C.~A.} \bibnamefont{Ullrich}},
  \bibinfo{author}{\bibfnamefont{U.~J.} \bibnamefont{Gossmann}}
  \bibnamefont{and} \bibinfo{author}{\bibfnamefont{E.~K.~U.}
  \bibnamefont{Gross}}, \bibinfo{journal}{Phys. Rev. Lett.}
  \textbf{\bibinfo{volume}{74}}, \bibinfo{pages}{872} (\bibinfo{year}{1995}).

  \bibitem[{\citenamefont{Ullrich et~al.}(1995)\citenamefont{Ullrich, Gossmann
  and Gross}}]{Ullrich1995b}
\bibinfo{author}{\bibfnamefont{C.~A.} \bibnamefont{Ullrich}},
  \bibinfo{author}{\bibfnamefont{U.~J.} \bibnamefont{Gossmann}}
  \bibnamefont{and} \bibinfo{author}{\bibfnamefont{E.~K.~U.}
  \bibnamefont{Gross}}, \bibinfo{journal}{Ber. Bunsenges. Phys. Chem.}
  \textbf{\bibinfo{volume}{99}}, \bibinfo{pages}{488-497} (\bibinfo{year}{1995}).

  \bibitem[{\citenamefont{van Leeuwen}(1996)}]{Leeuwen1996}
\bibinfo{author}{\bibfnamefont{R.}~\bibnamefont{van Leeuwen}},
  \bibinfo{journal}{Phys. Rev. Lett.} \textbf{\bibinfo{volume}{76}},
  \bibinfo{pages}{3610} (\bibinfo{year}{1996}).

   \bibitem[{\citenamefont{G{\"o}rling}(1997)}]{Gorling1997}
\bibinfo{author}{\bibfnamefont{A.}~\bibnamefont{G{\"o}rling}},
  \bibinfo{journal}{Phys. Rev. A} \textbf{\bibinfo{volume}{55}},
  \bibinfo{pages}{2630} (\bibinfo{year}{1997}).

\bibitem[{\citenamefont{K{\"u}mmel and Perdew}(2003)}]{Kummel2003}
\bibinfo{author}{\bibfnamefont{S.}~\bibnamefont{K{\"u}mmel}} \bibnamefont{and}
  \bibinfo{author}{\bibfnamefont{J.~P.} \bibnamefont{Perdew}},
  \bibinfo{journal}{Phys. Rev. Lett.} \textbf{\bibinfo{volume}{90}},
  \bibinfo{pages}{043004} (\bibinfo{year}{2003}).

\bibitem[{\citenamefont{K{\"u}mmel and Kronik}(2008)}]{Kummel2008}
\bibinfo{author}{\bibfnamefont{S.}~\bibnamefont{K{\"u}mmel}} \bibnamefont{and}
  \bibinfo{author}{\bibfnamefont{L.}~\bibnamefont{Kronik}},
  \bibinfo{journal}{Rev. Mod. Phys.} \textbf{\bibinfo{volume}{80}},
  \bibinfo{pages}{3} (\bibinfo{year}{2008}).

\bibitem[{\citenamefont{Shore and Knight}(1993)}]{Shore1993}
\bibinfo{author}{\bibfnamefont{B.~W.} \bibnamefont{Shore}} \bibnamefont{and}
  \bibinfo{author}{\bibfnamefont{P.~L.} \bibnamefont{Knight}},
  \bibinfo{journal}{J. Mod. Optic.} \textbf{\bibinfo{volume}{40}},
  \bibinfo{pages}{1195} (\bibinfo{year}{1993}).

\bibitem[{\citenamefont{Braak}(2011)}]{Braak2011}
\bibinfo{author}{\bibfnamefont{D.}~\bibnamefont{Braak}},
  \bibinfo{journal}{Phys. Rev. Lett.} \textbf{\bibinfo{volume}{107}},
  \bibinfo{pages}{100401} (\bibinfo{year}{2011}).

\bibitem[{\citenamefont{Wolf et~al.}(2013)\citenamefont{Wolf, Vallone, Romero,
  Kollar, Solano and Braak}}]{Wolf2013}
\bibinfo{author}{\bibfnamefont{F.~A.} \bibnamefont{Wolf}},
  \bibinfo{author}{\bibfnamefont{F.}~\bibnamefont{Vallone}},
  \bibinfo{author}{\bibfnamefont{G.}~\bibnamefont{Romero}},
  \bibinfo{author}{\bibfnamefont{M.}~\bibnamefont{Kollar}},
  \bibinfo{author}{\bibfnamefont{E.}~\bibnamefont{Solano}} \bibnamefont{and}
  \bibinfo{author}{\bibfnamefont{D.}~\bibnamefont{Braak}},
  \bibinfo{journal}{Phys. Rev. A} \textbf{\bibinfo{volume}{87}},
  \bibinfo{pages}{023835} (\bibinfo{year}{2013}).

\bibitem[{\citenamefont{Faisal}(1987)}]{Faisal}
\bibinfo{author}{\bibfnamefont{F.~H.} \bibnamefont{Faisal}},
  \emph{\bibinfo{title}{Theory of Multiphoton Processes}}
  (\bibinfo{publisher}{Springer}, \bibinfo{address}{Berlin},
  \bibinfo{year}{1987}).

\bibitem[{\citenamefont{Marques et~al.}(2010)\citenamefont{Marques, Ullrich,
  Nogueira, Rubio, Burke and Gross}}]{Marques}
\bibinfo{author}{\bibfnamefont{M.~A.~L.} \bibnamefont{Marques}},
  \bibinfo{author}{\bibfnamefont{C.~A.} \bibnamefont{Ullrich}},
  \bibinfo{author}{\bibfnamefont{F.}~\bibnamefont{Nogueira}},
  \bibinfo{author}{\bibfnamefont{A.}~\bibnamefont{Rubio}},
  \bibinfo{author}{\bibfnamefont{K.}~\bibnamefont{Burke}} \bibnamefont{and}
  \bibinfo{author}{\bibfnamefont{E.~K.~U.} \bibnamefont{Gross}},
  \emph{\bibinfo{title}{Time-Dependent Density-Functional Theory}}, vol.
  \bibinfo{volume}{706} (\bibinfo{publisher}{Springer},
  \bibinfo{address}{Berlin}, \bibinfo{year}{2010}).

\bibitem[{\citenamefont{Hedin}(1965)}]{Hedin1965}
\bibinfo{author}{\bibfnamefont{L.}~\bibnamefont{Hedin}},
  \bibinfo{journal}{Phys. Rev.} \textbf{\bibinfo{volume}{139}},
  \bibinfo{pages}{A796} (\bibinfo{year}{1965}).

\bibitem[{\citenamefont{Onida et~al.}(2002)\citenamefont{Onida, Reining and
  Rubio}}]{Onida2002}
\bibinfo{author}{\bibfnamefont{L.}~\bibnamefont{Onida}},
  \bibinfo{author}{\bibfnamefont{L.}~\bibnamefont{Reining}} \bibnamefont{and}
  \bibinfo{author}{\bibfnamefont{A.}~\bibnamefont{Rubio}},
  \bibinfo{journal}{Rev. Mod. Phys.} \textbf{\bibinfo{volume}{74}},
  \bibinfo{pages}{601} (\bibinfo{year}{2002}).

\bibitem[{\citenamefont{Streltsov et~al.}(2010)\citenamefont{Streltsov, Alon
  and Cederbaum}}]{Streltsov2010}
\bibinfo{author}{\bibfnamefont{A.~I.} \bibnamefont{Streltsov}},
  \bibinfo{author}{\bibfnamefont{O.~E.} \bibnamefont{Alon}} \bibnamefont{and}
  \bibinfo{author}{\bibfnamefont{L.~S.} \bibnamefont{Cederbaum}},
  \bibinfo{journal}{Phys. Rev. A} \textbf{\bibinfo{volume}{81}},
  \bibinfo{pages}{022124} (\bibinfo{year}{2010}).

\bibitem[{\citenamefont{Flick et~al.}(2014)\citenamefont{Flick, Appel and
  Rubio}}]{Flick2014}
\bibinfo{author}{\bibfnamefont{J.}~\bibnamefont{Flick}},
  \bibinfo{author}{\bibfnamefont{H.}~\bibnamefont{Appel}} \bibnamefont{and}
  \bibinfo{author}{\bibfnamefont{A.}~\bibnamefont{Rubio}}, \bibinfo{journal}{J.
  Chem. Theory Comput.} \textbf{\bibinfo{volume}{10}}, \bibinfo{pages}{1665}
  (\bibinfo{year}{2014}).


\bibitem[{\citenamefont{Wijewardane and Ullrich}(2008)}]{Wijewardane2008}
\bibinfo{author}{\bibfnamefont{H.~O.} \bibnamefont{Wijewardane}}
  \bibnamefont{and} \bibinfo{author}{\bibfnamefont{C.~A.}
  \bibnamefont{Ullrich}}, \bibinfo{journal}{Phys. Rev. Lett.}
  \textbf{\bibinfo{volume}{100}}, \bibinfo{pages}{056404}
  (\bibinfo{year}{2008}).

\bibitem[{\citenamefont{Linz}(1969)}]{Linz1969}
\bibinfo{author}{\bibfnamefont{P.}~\bibnamefont{Linz}},
  \bibinfo{journal}{Comput. J.} \textbf{\bibinfo{volume}{12 (4)}},
  \bibinfo{pages}{393} (\bibinfo{year}{1969}).

\bibitem[{\citenamefont{Castro et~al.}(2004)\citenamefont{Castro, Marques and
  Rubio}}]{Castro2004}
\bibinfo{author}{\bibfnamefont{A.}~\bibnamefont{Castro}},
  \bibinfo{author}{\bibfnamefont{M.~A.~L.} \bibnamefont{Marques}}
  \bibnamefont{and} \bibinfo{author}{\bibfnamefont{A.}~\bibnamefont{Rubio}},
  \bibinfo{journal}{J. Chem. Phys.} \textbf{\bibinfo{volume}{121}},
  \bibinfo{pages}{3425} (\bibinfo{year}{2004}).

\end{thebibliography}


\end{document}